# Coherent manipulation of a three-dimensional maximally entangled state

Shilong Liu,[1,2] Zhiyuan Zhou,[1,2,3,*] Shikai Liu,[1,2] Yinhai Li,[1,2] Yan Li,[1,2] Chen Yang,[1,2] Zhaohuai Xu,[1,2] Zhaodi Liu,[1,2] Guangcan Guo,[1,2] and Baosen Shi[1,2,3,†]

[1]*Key Laboratory of Quantum Information, University of Science and Technology of China, Hefei, Anhui 230026, China*
[2]*Synergetic Innovation Center of Quantum Information and Quantum Physics,
University of Science and Technology of China, Hefei, Anhui 230026, China*
[3]*Heilongjiang Provincial Key Laboratory of Quantum Regulation and Control, Wang Da-Heng Collaborative Innovation Center,
Harbin University of Science and Technology, Harbin 150080, China*

Maximally entangled photon pairs with a spatial degree of freedom is a potential way for realizing high-capacity quantum computing and communication. However, methods to generate such entangled states with high quality, high brightness, and good controllability are needed. Here, a scheme is experimentally demonstrated that generates spatially maximally entangled photon pairs with an orbital angular momentum degree of freedom via spontaneous parametric down-conversion in a nonlinear crystal. Compared with existing methods using postselection, the technique can directly modulate the spatial spectrum of down-converted photon pairs by engineering the input pump beam. In addition, the relative phase between spatially entangled photon pairs can be easily manipulated by preparing the relative phase of input pump states.

## I. INTRODUCTION

Engineering the entangled state is an important direction in quantum technology and forms a basis for quantum information processing. In particular, both the realization of a high-dimensional, maximally entangled state (MES) and its coherent manipulation are indispensable for investigating fundamental quantum physics, such as nonlocality [1,2]. Maximally entangled states are a fundamental resource for quantum communication protocols such as teleportation [3], storage [4,5], and secure cryptography [6,7]. Furthermore, applications to multimode quantum computation, such as high-dimensional quantum gates and Bell basis [8,9], will be important.

In photonic systems, a high-dimensional MES can be constructed by entangling two photons in a spatial degree of freedom [10–18] or in a degree of freedom of photon number [19], path [20,21], frequency [22], or temporal modes [23–26]. High-dimensional MES encoded in orbital angular momentum (OAM) via spontaneous parameter down-conversion (SPDC) has been a well-known and effective method [11,27], and much progress has been made [2,12–14]. Specifically, a three-dimensional MES of $|\varphi\rangle_d = 1/\sqrt{3}(|0, 0\rangle + |1, 1\rangle + |2, 2\rangle)$ was prepared in Ref. [13]. Recently, the nonlocality of a higher dimensional MES ($d = 12$) was studied in Ref. [2] using postselection of down-converted photons. In addition, a high-dimensional multiphoton entanglement was obtained with two nonlinear crystals [14].

However, most of OAM-based MES are generated via postselection of the quantum state ($|\varphi\rangle = \sum_{l=-\infty}^{\infty} c_l |l\rangle_A |-l\rangle_B$) produced by using SPDC with a strong Gaussian pump $E_{L=0}$. An example was reported in Ref. [28], which used Procrustean filtering. For simplicity, we use the notation of $|L\rangle_p$ to represent the pump beam $E_{L=L}$ with the only azimuthal variable; i.e., $|0\rangle_p$ describes the Gaussian beam as a pump.

However, the state generated by postfiltering is not indeed a maximally entangled state. This type of scheme has several disadvantages. On the one hand, because the amplitude $c_{|l|>0}$, determined by the spiral bandwidth (azimuthal Schmidt number) [29,30], is always less than $c_{|l|=0}$, many useful photon pairs with lower order modes (i.e., $|00\rangle$) are lost in the process of filtering. This is a more serious problem when constructing a higher dimensional MES. On the other hand, it is difficult to change the relative phase between different modes, while the phase between them is fundamental and important to increase the freedom of modulations for quantum information processes [8]. This is especially true for constructing high-dimensional Bell basis [9], symmetric or antisymmetric entangled state [16], and quantum dense coding [31]. Furthermore, to experimentally adjust the spiral bandwidth $c_l$, some parameters must be changed, such as input beam waists or the length of the nonlinear crystal [32], which is unwieldy in the experiment. Fortunately, utilization of a superposition state as the pump can overcome these shortcomings, and there are several theoretical protocols for engineering a high-dimensional MES [30,32,33]. For example, in 2003, Torner *et al.* proposed a scheme with the pump of a superposition OAM state to prepare a maximally entangled qu-quart $|00\rangle + |11\rangle + |22\rangle + |33\rangle$ [32]. However, due to the critical conditions, it is still challenging to generate such a state without postselection in an experiment. Recently, a spatial Bell state with a transverse Hermite-Gaussian mode has been directly generated without spatial filtration via SPDC [10]. Certainly, it is becoming a very promising direction to engineer high-dimensional entanglement by changing the profile of pump skillfully. Since the OAM degree of freedom of a photon is easier to manipulate and measure [34], it would be fascinating to generate an OAM-based MES without postselection.

---
[*]zyzhouphy@ustc.edu.cn
[†]drshi@ustc.edu.cn

Here, we first realize a two-photon, three-dimensional, OAM-based MES without postselection via the SPDC process and experimentally demonstrate the ability of amplitude and phase modulation in our system. The approach was to engineer the pump beam with an arbitrary superposition of Laguerre-Gauss (LG) modes. The state of down-converted photon pairs could then be manipulated independently based on OAM conservations in SPDC. To check the nonclassical characteristics of the prepared three-dimensional MES, we performed the Bell-type inequality ($S = 2.3729 \pm 0.0159$) and high-dimensional quantum-state tomography ($F = 0.8581 \pm 0.0028$). Also, the spatial spectrum of the down-converted photons was calculated. Comparing with the existing schemes, there were some advantages of these protocols: (i) It did not require postselections and (ii) the arbitrary phase between different superposition terms could be easily engineered by preparing the input beams. This work provides a convenient and efficient platform to explore potential applications in quantum communications.

## II. PRINCIPLE

The spatially entangled down-converted photon pairs in SPDC can be written as $\sum_{l_s, p_s} \sum_{l_i, p_i} c_{p_s, p_i}^{l_s, l_i} |l_s, p_s\rangle |l_i, p_i\rangle$ in the Laguerre-Gauss mode basis [29], where the indices of $l$ and $p$ described the azimuthal and radial OAM modes; $|c_{p_s, p_i}^{l_s, l_i}|^2$ represents a coincidence probability for finding one signal photon in spatial modes of $l_s$, $p_s$ and one idler photon in spatial modes of $l_i$, $p_i$. The distributions of coincidence probability $\{|c_{p_s, p_i}^{l_s, l_i}|^2\}$ are named as spatial spectrum distribution for simplicity. Because the radial value of $p$ between pumps and down-converted photon pairs are not conserved [30], we only consider the azimuthal values of $l$, and the full spatial spectrum distribution both $p$ and $l$ for nonthin nonlinear crystal are still analyzed in supplementary materials. Down-converted photon pairs have the form $|\varphi\rangle_{\text{Single}} = \sum_{l=-\infty}^{\infty} c_l |l\rangle_A |L-l\rangle_B$ for a single pure-state pump $|L\rangle_p$; by analogy, it can be written as $|\varphi\rangle_{\text{Multi}} = \sum_L \sum_{l=-\infty}^{\infty} c'_l |l\rangle_A |L-l\rangle_B$ when the pump is a superposition of multiple LG modes $\sum_L C_L |L\rangle_p$ ($C_L$ is a complex number, i.e., $C_L = |C_L|e^{i\theta_L}$). Engineering a suitable pump state of $C_{-2}|-2\rangle_p + C_0|0\rangle_p + C_2|2\rangle_p$, the down-converted entangled state can be generated as following based on OAM conservation($l_p = l_s + l_i$): $|\varphi\rangle_{\text{MES}} = \xi_{\text{MES}}(|-1-1\rangle + |00\rangle + |11\rangle + \sum_L \sum_{l=-\infty}^{\infty} c'_l \xi_L^l |l\rangle |L-l\rangle)$, where the $\xi_L^l$ are the amplitudes of high-order OAM entangled pairs, which can be suppressed by changing the pump beam and signal-idler beam widths [29,33]. Then the state is reduced to a three-dimensional MES in the subspace.

Figure 1 depicts the main principle of the scheme, including the spatial spectrum distributions and the corresponding intensity profiles of the input pumps. The effective spatial spectrum is distributed along the corresponding diagonal elements based on the OAM conserved. For example, the green rectangular and red boxes in Fig. 1(e) are the situations for the pump with $|L\rangle_p = |0\rangle_p, |2\rangle_p, |-2\rangle_p$, respectively. If the pump is a superposition state, i.e., $|L\rangle_p = 1/\sqrt{3}(|0\rangle_p + |2\rangle_p + |-2\rangle_p)$, the output spatial spectrum distribution equals the effects of a linear superposition for each pure-state pump [33], and the result is presented in Fig. 1(a). By increasing the occupations of pump $|\pm 2\rangle_p$, the two smaller heights of $|1\rangle_s |1\rangle_i$ and $|-1\rangle_s |-1\rangle_i$ can be grown from Figs. 1(a) to 1(b). Thus, an arbitrary three-dimensional entangled state can be generated. For an extreme situation with only two pump vortex pumps $|\pm 2\rangle_p$, the output is a two-dimensional MES [see Fig. 1(d)], which has been widely used in quantum information [35]. Figure 1(c) shows the MES of $|\varphi\rangle_{\text{MES}} = 1/\sqrt{3}(|-1\rangle_s|-1\rangle_i + |0\rangle_s|0\rangle_i + |1\rangle_s|1\rangle_i)$, where the high-order component can be reduced by modulating the beam waist of the pump and optimizing overlaps between distributions of photon pairs and phase holograms in a spatial light modulator (SLM) [29,33]. Figures 1(f)–1(h) indicate the required intensity profiles of the input pump for Figs. 1(b)–1(d), respectively.

In addition, the relative phase between the pump superposition state can be fully transferred to the corresponding OAM entangled photon pairs. For example, the output would be $|\varphi\rangle_{\text{MES}} = 1/\sqrt{3}(|-1\rangle_s|-1\rangle_i + e^{i\theta_0}|0\rangle_s|0\rangle_i + e^{i\theta_2}|1\rangle_s|1\rangle_i)$ under the input pump of $N(\sqrt{2.5}|-2\rangle_p + e^{i\theta_0}|0\rangle_p + \sqrt{2.5}e^{i\theta_2}|2\rangle_p)$.

Figure 2 shows the optical paths for generating and coherently modulating a three-dimensional MES. The 780-nm input source was a Ti:sapphire continuous laser (MBR110, Coherent). A visible-wavelength spatial light modulator (SLM-V, 512 × 512 SLM, model P512-0785, Meadowlark, Optics) was used to modulate the input laser beams. The laser beam carrying the acquired phase in SLM-V was exactly imaged on the center of nonlinear crystals (PPKTP, 10 mm) by a lens ($f = 100$ mm). The beam waist in the center of the crystal was 31 $\mu$m. More specially, first, the generated infrared OAM-entangled 1550-nm photon was long-pass filtered. Then, it passed through an imaging system with an infrared lens ($f = 150$ mm) and a polarization beam splitter (PBS) before reaching the plane of SLM-I (PLUTO-2-TELCO-013, 1920 × 1080 pixels). Finally, by a group of coupling lenses, the entangled photons in SLM-I without vortex phase information are imaged on the surfaces of fibers, which are ported to superconducting nanowire single-photon detectors (SSPDs, SCONTEL) that had four channels with the ports of single-mode fiber for detections. The imprinted phase profiles in both SLMs were given by amplitude-encoded phase-only holograms [see Appendix B]. In this type of encoded hologram, adjustment of the beam waists for different OAM modes was required for measurements of optimal mode overlaps.

## III. RESULTS

Results for engineering a three-dimensional arbitrary entangled state are presented in Fig. 3. Two manipulation steps were employed. One was to change the real part of the coefficients $|C_j|$ for three inputs, named as amplitude modulations. The corresponding spatial spectrum distributions are presented in the Appendixes. Figure 3(a) shows the situation of three-dimensional MES, where the $x(y)$ axis was the topologic charge of pure-OAM states loaded on the ports of A(B) in the SLM-I (Fig. 2). The $z$ axis represented the measured coincidence per second. To evaluate the spectral brightness of the infrared photonic pairs, the measurement time was up to 20 s. Accounting for all the losses of collections $\alpha_A$, $\alpha_B$ and detections $\eta_A$, $\eta_B$, the inferred spectral brightness $B_L = N_c/\alpha_A\alpha_B\eta_A\eta_B P \Delta\lambda$ for the maximum coincidence, i.e., $|00\rangle$, was approximately $3.0 \times 10^4$/(s mW nm) and $1.0 \times 10^4$/(s mW nm) for the single pure state pump with $L = 0$ and $\pm 2$, respectively, where $\Delta\lambda$ (2.4 nm) was the spectral bandwidth of the signal and idler photons [36]. The overall losses, named as the transmission losses, $\alpha_A\eta_A$, $\alpha_B\eta_B$ of each photon pair from generation to SSPDs were 10% and 4% for pumps with $L = 0$ and $\pm 2$.

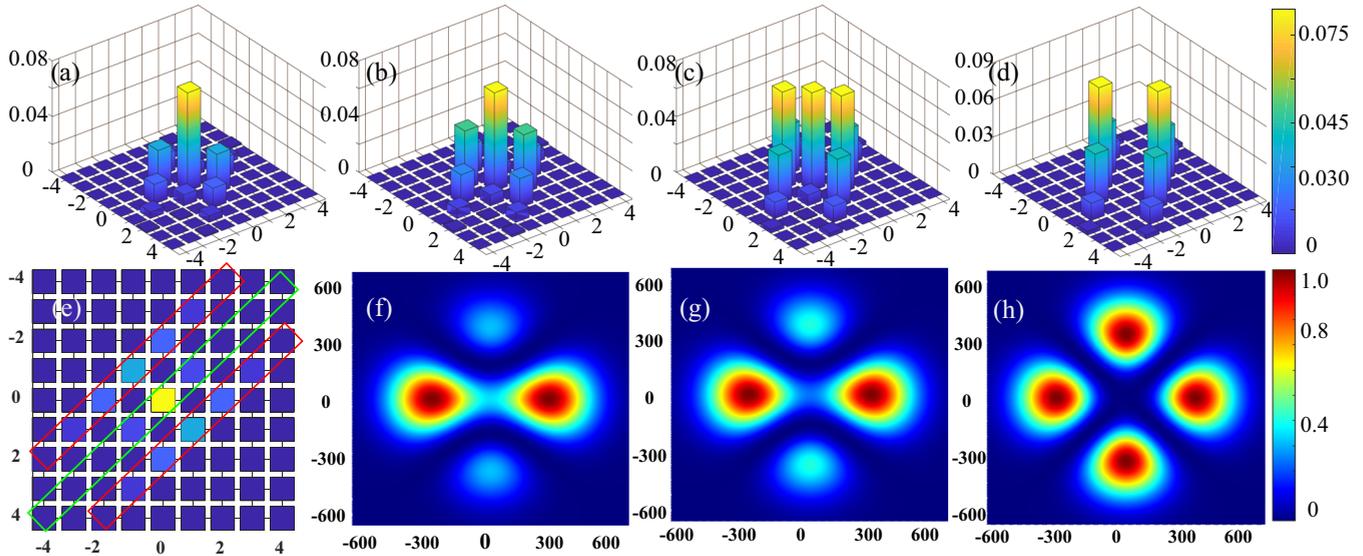

FIG. 1. Theoretical results of spatial spectrum distributions of down-converted photons and the required intensity profiles of the input beams. [(a)–(d)] The spatial spectrum distributions for the input pump $|-2\rangle_p + |0\rangle_p + |2\rangle_p$, $\sqrt{1.5}|-2\rangle_p + |0\rangle_p + \sqrt{1.5}|2\rangle_p$, $\sqrt{2.5}|-2\rangle_p + |0\rangle_p + \sqrt{2.5}|2\rangle_p$, $|-2\rangle_p + |2\rangle_p$, respectively, where the beam widths of the pump and detection are $w_p = 1.0$, $\gamma_s = \gamma_i = 0.5$ [29,33]. Actually, the coefficients of input state for MES need to change based on the different system of collections. (e) Two-dimnensional (2D) perspective of Fig. 1(a), where the $x(y)$ labels represent the modes of signal and idler photons. [(f)–(h)] Corresponding required intensity profiles, where the wavelength of the pump is 780 nm.

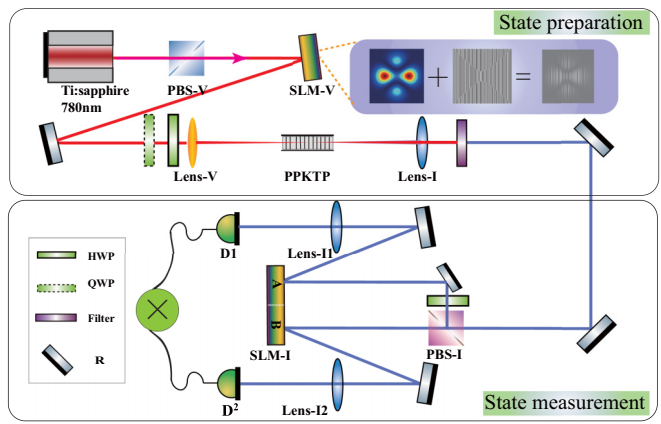

FIG. 2. Optical layout for generating and detecting maximally entangled states. SLM-V and SLM-I are a spatial visible and infrared light modulator used to modulate and measure the OAM-entangled state. The state $|-l\rangle_s |-l\rangle_i$ could be converted into $|-l\rangle_s |l\rangle_i$ by changing the number of reflection mirrors in the idler path.

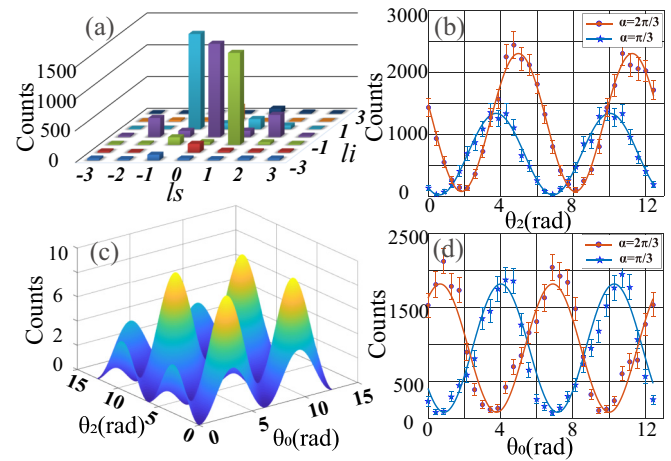

FIG. 3. Spatial spectral distributions and interference results. (a) Coincidences for three-dimensional MES per second. [(b), (d)] Coincidence and interference curve fits for various phase angle of $1/\sqrt{3}(|-2\rangle_p + e^{i\theta_0}|0\rangle_p + e^{i\theta_2}|2\rangle_p)$, where the pentagram and circle data respectively represent two phase parameters $\alpha = 2\pi/3$ and $\alpha = \pi/3$ in SLM-I. The error bars were $\pm 2$ standard deviations, assuming the data followed Poisson's distribution. Each coincidence was recorded over 20s. (c) Theoretical interferences of MES with two phase factors.

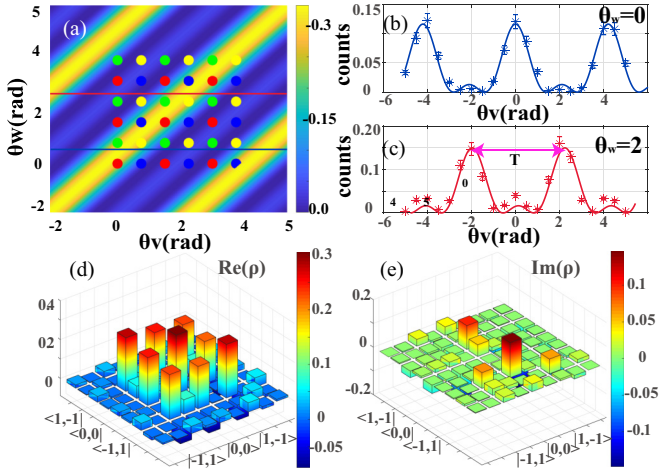

FIG. 4. Results of Bell inequality and tomographic reconstructions. [(a), (c)] Theoretical Bell-type 3-D surface and measured interference; the period $T$ in panel (c) was used to ensure the spacing of the measurement basis. [(d), (e)] The real and imaginary parts of density matrices of $d = 3$, respectively. There were two relatively high (around 0.1) imaginary parts in panel (b), which was the result of a constant phase between low- and high-order states. Each coincidence for quantum-state tomography was recorded over 20 s.

The other manipulation was to change the phase of the coefficients $|C_j|e^{i\theta_j}$ for input states, named as phase modulations. To prove that this phase was fully converted from input states to quantum entangled states, the measurement basis $|\varphi\rangle_A \otimes |\varphi\rangle_B (= N(|-1\rangle_A + e^{i\alpha}|0\rangle_A + e^{i\alpha}|1\rangle_A) \otimes (|-1\rangle_B + e^{-i\alpha}|0\rangle_B + e^{-i\alpha}|1\rangle_B))$ was designed and loaded on SLM-I. Then, the coincidences $|\langle\varphi|_A \langle\varphi|_B |\varphi\rangle_{\text{MES}}|^2$ were derived as $N\{3 + 2[\cos(\alpha + \theta_0) + \cos(2\alpha + \theta_2) + \cos(\alpha + \theta_2 - \theta_0)]\}$. Figure 3(c) exhibits the three-dimensional (3-D) surface with variations of the two phase angles. Figures 3(b) and 3(d) present two special interference curves for $\theta_2 = 0 - 3\pi, \theta_0 = \pi/4$ and $\theta_0 = 0 - 3\pi, \theta_2 = 0$. For each of the subfigures, the parameters $\alpha$ were $2\pi/3$ and $\pi/3$, respectively, and the solid lines were fits based on theoretical predictions. The visibilities ($V = C_{\text{Max}} - C_{\text{Min}}/C_{\text{Max}} + C_{\text{Min}}$) of the red and blue solid curves were $0.9387 \pm 0.0178, 0.9735 \pm 0.0133$ for Fig. 3(b) and $0.9126 \pm 0.0382, 0.9116 \pm 0.0347$ for Fig. 3(d), respectively. In Fig. 3(d), the initial coincidence was not a minimum value, which illustrated that, for the prepared MES, there existed a basic constant phase factor between OAM entangled photon pairs.

To test the nonclassical characteristics of the OAM-based MES, measurement of the high-dimensional Bell inequalities was performed [1]. Many groups have successfully demonstrated the violation of the high-dimensional Bell inequality of OAM-based entangled photon pairs [2,4,13,35]. Here, the measurement basis $|\theta_{A,B}^{a,b}\rangle$ on SLM-I was designed as a three-dimensional superposition state [1,2]. Details are given in the Appendixes. For a three-dimensional MES, the coincidence formed a 3-D surface [Fig. 4(a)] with the variation of two angles. To test the Bell inequality, 36 measurements were divided into four groups four colors in Fig. 4(a) of nine samplings.

For a 10-mm-long type-II periodically poled potassium titanyl phosphate (PPKTP), the spatial parameters $\tau_a, \tau_b$ associated with the spacing of the measurement basis $|\theta_{A,B}^{a,b}\rangle$ were $4.2/2\pi$, by fitting the Bell-type interference curves in Figs. 4(b) and 4(c). By assuming that the coincidences followed Poissons distribution, the Bell inequality $S_3 = 2.3735 \pm 0.0159$ in $3 \times 3$ dimensions, which was more than 24 standard deviations. This violation of a Bell inequality directly indicated that the generated state was entangled.

In the next, density matrices of three-dimensional MES were constructed via high-dimensional quantum-state tomography (QST) via mutually unbiased measurements. In the experiment, we employ the method of projective measurements that include an SLM, single-mode fibers, and coincidence-counting electronics, where the SLM is used to erase the phase information of photons; the single-mode fiber ensures that the photon with the only Gaussian profile can be coupled to detectors. The details can be found in Refs. [37–39]. The corresponding reconstructed density matrices were $\rho = N \sum_{u,v,j,k=1}^{d^2} (A_{uv}^{jk})^{-1} n_{uv} \lambda_j \otimes \lambda_k$. Details on parameters are described in the Appendixes. To ensure that the density matrices were physical, i.e., had the property of positive semidefiniteness [40], maximum likelihood estimation methods were performed during the reconstructions. The reconstructed density matrices are shown in Fig. 4 with a 3D perspective, where Figs. 4(d) and 4(e) represented the real and imaginary parts, respectively. The fidelity $F = \text{Tr}[\sqrt{\sqrt{\rho}\rho_{\text{exp}}\sqrt{\rho}}]^2$ was $0.8581 \pm 0.0028$, where $0.0028$ was the standard deviation obtained by statistical simulations that assumed that each photonic coincidence followed Poisson's distribution. There were two relatively high imaginary parts in Fig. 4(e), which was the result of a constant phase between low-order ($|00\rangle$) and high-order ($|1-1\rangle$) entangled states. The constant phase could be concluded by the interference curves in Fig. 3(d). Furthermore, calculations of the linear entropy $S_{\text{ent}} = 1 - \text{Tr}(\rho_{\text{exp}}^2)$ yielded $S_{\text{ent}} = 0.2451 \pm 0.0174$. These measurement parameters of MES had a little gap in contrast to the theoretical ideal, which were attributed to imperfect mode overlaps between distributions of photons and the phase hologram in SLM-I, and cross talk between different OAM modes. Nevertheless, the main MES features exhibited here were close to other schemes [13,37]. Hence, the method was reliable and practical.

## IV. DISCUSSION

An effective scheme to engineer a three-dimensional MES with arbitrary relative phase was demonstrated experimentally. The results were in good agreement with theoretical predictions. A well-controlled and high-quality source could thus provide a platform for exploring the multimode quantum information. Recently, we realized a OAM-based Schrödinger cat state $|\alpha\rangle = \sum_{L=0}^{\infty} c_l |L\rangle$ with plenty of OAM modes [41], which ensures that we can explore the high-dimensional MESs under the arbitrary and complex input OAM superposition state. The scheme to prepare an infrared biphoton three-dimensional MES had some advantages compared with other techniques. An arbitrary modulation could be made in amplitude and phase, which could be an attractive and potential platform for multimode quantum information processing. For this method, the brightness of the high-dimensional MES depended on the pump generation rates of high-order OAM modes. For example, the brightness of $|1\rangle_s |1\rangle_i$ was $1.0 \times 10^4$/(s mW nm) for the pump $|2\rangle_p$. The brightness will naturally decrease with the increase of the topologic charges, which is a bottleneck restricting the increase in dimensions. In addition, it would be an interesting and valuable challenge to decrease the noise component of high orders or to decrease the spiral bandwidth (Schmidt number) when the single high-order pure state acts as a pump. One solution would be modifications of the input and collection parameters, but that will lead to decreases in source brightness [33,42]. Another solution would be to increase the spatial Schmidt numbers (spatial purity) by modulating the frames of nonlinear crystals. In the frequency domain, much progress has been made by modulating the poling period or duty cycles [43–45]. However, in the spatial domain, there has been little experimental progress.

*Note added.* We note a closely related work [46].

## ACKNOWLEDGMENTS

We thanks Dr. Huan Cao (University of Science and Technology of China) for providing a spatial light modulator. This work is supported by the Anhui Initiative in Quantum Information Technologies (Grant No. AHY020200); National Natural Science Foundation of China (Grants No. 61435011, No. 61525504, and No. 61605194); China Postdoctoral Science Foundation (Grants No. 2016M590570 and No. 2017M622003); and Fundamental Research Funds for the Central Universities.

## APPENDIX A: SPONTANEOUS PARAMETRIC DOWN CONVERSION (SPDC) AND SPIRAL BANDWIDTH (SB)

With a nonlinear crystal, collinear down conversion, and illuminated by Laguerre-Gaussian (LG) modes, the down-converted photon pairs are given by [29,30]

$$|\Phi\rangle = \sum_{l_s,p_s} \sum_{l_i,p_i} c^{l_s,l_i}_{p_s,p_i} |l_s, p_s\rangle |l_i, p_i\rangle, \quad (A1)$$

where $c^{l_s,l_i}_{p_s,p_i}$ describes the amplitude of the down-converted photon pairs. It can be calculated from the overlap integral under the LG mode pump:

$$c^{l_s,l_i}_{p_s,p_i} = \langle l_s, p_s; l_i, p_i | |\Phi\rangle = \int_{-L/2}^{L/2} \int_0^{2\pi} d\phi \int_0^\infty \\ \times r \, dr \, LG^{l_p}_p \left[LG^{l_s}_{p_s}(r,\phi)\right]^\dagger \left[LG^{l_i}_{p_i}(r,\phi)\right]^\dagger dz, \quad (A2)$$

where the pump light is the LG modes with the following form in the cylindrical coordinate [27]:

$$|L\rangle = E_L = \sqrt{\frac{2}{\pi |L|!}} \frac{1}{w(z)} \left[\frac{\sqrt{2}r}{w(z)}\right]^{|L|} \exp\left[-\frac{r^2}{w(z)^2}\right] \\ \times \exp\left\{i\left[L\phi - \frac{kr^2 z}{2(z^2+z_r^2)} + (|L|+1)a\tan\left(\frac{z}{z_r}\right)\right]\right\}. \quad (A3)$$

The integral over the azimuthal coordinate $\int_0^{2\pi} d\phi \exp[i(l_p - l_s - l_i)\phi] = 2\pi \delta_{l_p, l_s+l_i}$ predicts the well-known OAM conservation. Substituting the LG mdoes into Eqs. (A1) and (A2), we obtain the amplitudes

$$c^{l_s,l_i}_{p_s,p_i} \\ \approx c(w_p, l_p, p_p, w_s, l_s, p_s, w_i, l_i, p_i, |L_c|) = \int_{-L_c/2}^{L_c/2} \\ \sqrt{2p_p! p_s! p_i! \pi^{-3} 2^{\sigma_l}(p_p + |l_p|)!(p_s + |l_s|)!(p_i + |l_i|)!} \\ \times \sum_{k=0}^{p_p} \sum_{i=0}^{p_i} \sum_{s=0}^{p_s} \frac{(-2)^{\sigma_t}(\sigma_t + \sigma_l/2)!}{(|l_p|+k)!(p_p-k)!(p_i-i)!(|l_i|+i)!} \\ \times \frac{(k! i! s!)^{-1}(w_p^{-2} + w_i^{-2} + w_s^{-2})^{-(\sigma_t+\sigma_l/2+1)}}{(|l_s|+s)!(p_s-s)! w_p^{2k+|l_p|+1} w_i^{2i+|l_i|+1} w_s^{2s+|l_s|+1}}, \quad (A4)$$

where $L_c$ is the length of nonlinear crystal; $\sigma_t (= k+i+s)$ and $\sigma_l (= |l_p|+|l_s|+|l_i|)$ are two abbreviated parameters; the waists for three beams are defined as $w(z) = w_0\sqrt{1+(z/f)^2}$; and $f$ is the Rayleigh distance $f = \pi w_0^2/\lambda$. The distributions of coincidence probability $c^{l_s,l_i}_{p_s,p_i}$, named the spatial spectrum distribution, can be simulated based on the parameters of pumps and signal-idler photons. A detailed descripton can also be found in Ref. [30].

## APPENDIX B: THE DETAILS OF MEASUREMENTS AND PHASE MODULATIONS

The pump light is a Laguerre-Gaussian (LG) mode with Eq. (A3), where $w(z)$ represents the beam's radius at the transmission distance $z$. When $z=0$, the beam waist is $w_0$; $L$ represents the topologic charge of vortex light. As the topologic charge $L$ is larger, the size of the beam waist increases. For generating a single pure OAM state, we load an amplitude-encoded phase-only hologram in the SLM, which has the following form:

$$\Phi(x, y)_{\text{holo}} = \text{sinc}(\pi M - \pi)\text{Arg}(E_L)\Lambda(x, y), \quad (B1)$$

where $M$ ($0 < M < 1$) is a normalized bound for the positive amplitude function, $E_L$ is an analytical function describing the pump, and $\Lambda(x, y)$ is an optical blazing phase for generating pure OAM quantum states. For generating a superposition state, we need to replace $E_L$ by $\sum E_L$. The phase encoded in another SLM for detections is the same as that above. For realizing an optimal overlap between measured phase and actually entangled photonic distribution, we design a feasible magnification of 13.6 by a lens placed after nonlinear crystals. Furthermore, we increase the spatial purity by adjusting the waist of the beam for different OAM modes [47]:

$$w_L(z=0) = w_0 \frac{2^{-|L|}(2|L|+1)!!}{|L|!}, \quad (B2)$$

where $n!! = 1 \times 3 \times 5 \times \cdots \times n$. Using this modulation, the area of phase for various mode will tend to equal area [47]. Figure 5 shows the acquired phase for a three-dimensional MES and the various spectrum distributions under the different pumps. Using the amplitude-modulation technology, we can engineer the output state from one to three dimensions. In sections of phase modulations, we find a basic offset phase existed between the superposition terms. Considering the following MES generated in SPDC,

$$\psi = \frac{1}{\sqrt{3}}(|-1\rangle_A |1\rangle_B + e^{i\theta_0} |0\rangle_A |0\rangle_B + e^{i\theta_2} |1\rangle_A |-1\rangle_B),$$

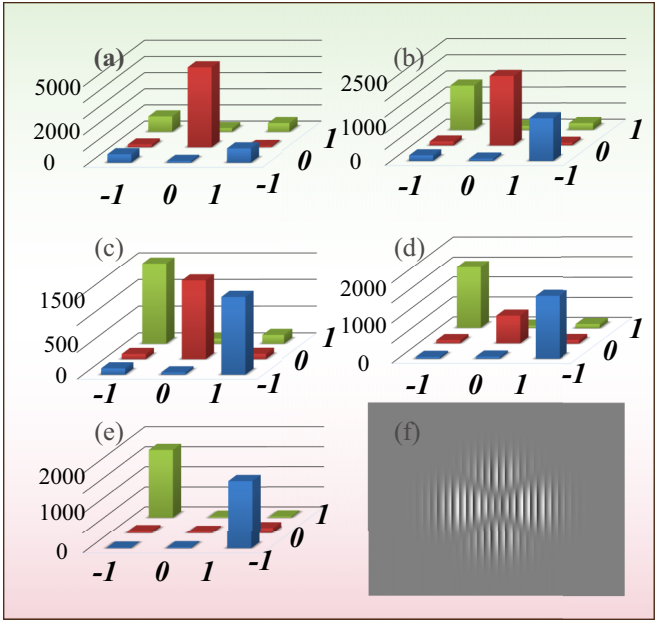

FIG. 5. Spatial spectrum distributions. [(a)–(e)] The coincidences for various amplitude-modulated input states as $|-2\rangle_p + |0\rangle_p + |2\rangle_p$, $2|-2\rangle_p + |0\rangle_p + 2|2\rangle_p$, $3|-2\rangle_p + |0\rangle_p + 3|2\rangle_p$, $4|-2\rangle_p + |0\rangle_p + 4|2\rangle_p$, and $|-2\rangle_p + |2\rangle_p$, respectively, where the normalized coefficients are ignored for simplicity. The coincidence times and windows are 1 s and 1.6 ns in the process of measurements. (f) The phase hologram in the SLM for generating a three-dimensional MES.

(B3)

where we ignore the global phase factor and the phases of $e^{(i\theta_0)}$ and $e^{(i\theta_2)}$ are relative to the first OAM mode $|-11\rangle$. For ensuring the detailed values, we design a measurement basic state:

$$|\psi_A\rangle \otimes |\psi_B\rangle = N(|-1\rangle_A + e^{i\alpha} |0\rangle_A + e^{i\alpha} |1\rangle_A)$$
$$\otimes (|-1\rangle_B + e^{-i\alpha} |0\rangle_B + e^{-i\alpha} |1\rangle_B). \quad \text{(B4)}$$

The coincidence can be inferred as

$$C(\theta_0, \theta_2) = |\langle\psi|_A \langle\psi|_B |\psi\rangle|^2 \approx 3 + 2[\cos(\alpha + \theta_0) + \cos(2\alpha + \theta_2) + \cos(\alpha + \theta_2 + \theta_0)], \quad \text{(B5)}$$

where $\theta_0$ and $\theta_2$ are two independent phase factors, and $C(\theta_0, \theta_2)$ forms a 3-D surface presented in Fig. 3(c) of the main text. By measuring the fitting lines in Fig. 3(b), the offset phase angle is determined to be around 0.6 rad between them.

## APPENDIX C: THE INTERFERENCE AND BELL INEQUALITY

The high-dimensional Bell inequality can be defined as [1]

$$S_d = \sum_{k=0}^{[d/2]-1} S_k = \sum_{k=0}^{[d/2]-1} \left(1 - \frac{2k}{d-1}\right)$$
$$\times \{[P(A_0 = B_0 + k) + P(B_0 = A_1 + k + 1)$$
$$+ P(A_1 = B_1 + k) + P(B_1 = A_0 + k)]$$
$$- [P(A_0 = B_0 - k - 1) + P(B_0 = A_1 - k)$$
$$+ P(A_1 = B_1 - k - 1) + P(B_1 = A_0 - k - 1)]\},$$

(C1)

where $P(A_a = i, B_b = j)$ is the normalized value associating the photonic coincidence $P(A_a = i, B_b = j) = C(A_a = i, B_b = j)/\sum_{j'i'=0}^{d-1} C(A_a = i', B_b = j')$. Each of coincidence can be written as $C(\theta_A, \theta_B) = |\langle\theta_A|\langle\theta_B||\psi_{\text{MES}}\rangle|^2$. Following by definitions in Ref. [2], two measurement angular states have such forms:

$$|\theta_{A,B}^{a,b}\rangle = \frac{1}{\sqrt{d}} \sum_{l=-[d/2]}^{[d/2]} \exp(i\theta_{A,B}^{a,b} g(l)) |l\rangle. \quad \text{(C2)}$$

All the meanings of parameters are the same as in Ref. [2], the only difference is the two angular parameters $\theta_A^a$, $\theta_B^b$:

$$\theta_A^a = \frac{2\pi}{d}\tau_a(v + a/2); \quad \theta_B^b = \frac{2\pi}{d}\tau_b[-w + (-1)^b/4].$$

(C3)

Because the nonlinear crystal used in our setup is a 10-mm long periodically poled potassium titanyl phosphate (PPKTP), there exists a parameter $\tau_a, \tau_b$, which can be fitted by Bell-type interference curves in Fig. 3(b). Considering the two angles are dependent continuous values, the coincidences forms a 3-D surface shown in Fig. 4(a) of the main text.

## APPENDIX D: THE DETAILS OF QUANTUM STATE TOMOGRAPHY

The theoretical density matrices of MES can be written as

$$\rho = |\psi\rangle_{\text{MES}} \otimes \langle\psi|_{\text{MES}}, \quad \text{(D1)}$$

where the MES is defined as $|\psi\rangle = 1/\sqrt{3} \cdot |-11\rangle + |00\rangle + |1-1\rangle$; each of OAM eigenstates can be expressed as a vector in a three-dimensional space, i.e., $|-1\rangle = [100]^T; |0\rangle = [010]^T; |1\rangle = [001]^T$. Using this definition, we can calculate the theoretical matrices of MES [37].

Experimentally, the density matrices of a three-dimensional MES can be reconstructed by high-dimensional quantum state tomography via mutually unbiased measurements (MUB). The corresponding reconstructed density matrices can be written as

$$\rho = N \sum_{u,v,j,k=1}^{d^2} (A_{uv}^{jk})^{-1} n_{uv} \lambda_j \otimes \lambda_k, \quad \text{(D2)}$$

where $N$ is the normalized coefficient; $A_{uv}^{jk}(=\langle\Psi_{uv}|\lambda_j \otimes \lambda_k|\Psi_{uv}\rangle)$ is the constant matrix associating with the fundamental matrix $\lambda_{j,k}$ and measurement basis; $|\Psi_{uv}\rangle$ and $(A_{uv}^{jk})^{-1}$ are the corresponding inverted matrices; $|\lambda_{j,k}\rangle$ is the elementary matrix associating SU(3) groups [39]; $|\Psi_{u,v}\rangle = |\Psi_u\rangle_A |\Psi_v\rangle_B^\dagger$ represents the measurement basis of signal (A) and idler photons (B); and $n_{uv} = Ntr(\Pi_{A,B}\rho_{\exp})$ is the coincidence photon number measured by electronic systems with the operations for ports A and B [37]. Based on the MUBs, each of measurement basis $|\Psi\rangle$ is a superposition state of the basic OAM eigenstates: $|\Psi\rangle = \sum_{L=[d/2]}^{[d/2]} c_{u,v,L} |L\rangle$, where coefficients can be calculated by the discrete Fourier transforms [37,38].

In principle, the density matrix can be reconstructed by Eq. (D2). The reconstructed density matrix may not be a physical density matrix; i.e., it has the property of positive semidefiniteness [40]. For overcoming this disadvantage, the maximum likelihood estimation method is used during the process of reconstructions. We build the likelihood function

$$L(t_1, t_2, \ldots, t_{81}) = \sum_{j=1}^{81} \frac{[N(\langle\Psi_j|\rho_{\exp}|\Psi\rangle_j - n_j]^2}{2N(\langle\Psi_j|\rho_{\exp}|\Psi\rangle_j}, \quad \text{(D3)}$$

where the $\rho_{\exp}$ has the same meaning as the formula in Ref. [40].